\documentclass[conference]{IEEEtran}
\IEEEoverridecommandlockouts
\usepackage{cite}
\usepackage{amsmath,amssymb,amsfonts}
\usepackage{algorithmic}
\usepackage{graphicx}
\usepackage{textcomp}
\usepackage{xcolor}
\usepackage{comment}
\usepackage{hyperref}
\usepackage{todonotes}
\def\BibTeX{{\rm B\kern-.05em{\sc i\kern-.025em b}\kern-.08em
    T\kern-.1667em\lower.7ex\hbox{E}\kern-.125emX}}

\newcommand{\brazil}{Brazil}
    
\begin{document}

\title{A Socio-Technical Grounded Theory on the Effect of Cognitive Dysfunctions in the Performance of Software Developers with ADHD and Autism}

\author{\IEEEauthorblockN{Kiev Gama}
\IEEEauthorblockA{\textit{Federal University of Pernambuco (UFPE)} \\
Recife, Brazil \\
kiev@cin.ufpe.br}
\and
\IEEEauthorblockN{Grischa Liebel}
\IEEEauthorblockA{\textit{Reykjavik University} \\
Reykjavik, Iceland \\
grischal@ru.is}
\and
\IEEEauthorblockN{Miguel Goulão}
\IEEEauthorblockA{\textit{NOVA University Lisbon} \\
Lisbon, Portugal \\
mgoul@fct.unl.pt}
\and
\IEEEauthorblockN{Aline Lacerda}
\IEEEauthorblockA{\textit{Federal University of Pernambuco (UFPE)} \\
Recife, Brazil \\
aline.lacerda@ufpe.br}
\and
\IEEEauthorblockN{Cristiana Lacerda}
\IEEEauthorblockA{\textit{Federal University of Pernambuco (UFPE)} \\
Recife, Brazil \\
cllp@cin.ufpe.br}
}
\maketitle

\begin{abstract}
The concept of neurodiversity, encompassing conditions such as Autism Spectrum Disorder (ASD), Attention-Deficit/Hyperactivity Disorder (ADHD), dyslexia, and dyspraxia, challenges traditional views of these neurodevelopmental variations as disorders and instead frames them as natural cognitive differences that contribute to unique ways of thinking and problem-solving. Within the software development industry, known for its emphasis on innovation, there is growing recognition of the value neurodivergent individuals bring to technical teams. Despite this, research on the contributions of neurodivergent individuals in Software Engineering (SE) remains limited. This interdisciplinary Socio-Technical Grounded Theory study addresses this gap by exploring the experiences of neurodivergent software engineers with ASD and ADHD, examining the cognitive and emotional challenges they face in software teams. Based on interviews and a survey with 25 neurodivergent and 5 neurotypical individuals, our theory describes how neurodivergent cognitive dysfunctions affect SE performance, and how the individuals' individual journey and various accommodations can regulate this effect. We conclude our paper with a list of inclusive Agile practices, allowing organizations to better support neurodivergent employees and fully leverage their capabilities.
\end{abstract}


\section{Introduction}

The concept of neurodiversity is an umbrella term for a range of neurocognitive developmental disorders (e.g., Autism Spectrum Disorder, Attention-Deficit/Hyperactivity Disorder, dyslexia, and dyspraxia) has gained increasing attention in recent years and challenges traditional perspectives on conditions~\cite{doyle2021diamond}. This new perspective encourages viewing these conditions not as disorders but as natural variations in cognitive functioning. It posits that these differences, resulting from distinctive brain structures and processes (i.e., neurodivergent), lead to unique ways of thinking and behaving~\cite{sonugabarke2021neurodiversity}.

The software development industry, known for its emphasis on innovation and problem-solving, has started to recognize the potential of fostering diversity in all its forms, including neurodiversity. While much of the focus in diversity efforts has centered on gender, ethnicity, and race, there is an emerging awareness of the value neurodivergent individuals bring to technical teams. Research exploring the contributions of neurodivergent individuals, however, remains limited in Software Engineering (SE)~\cite{rodriguez2021perceived}. Neurodivergent individuals, including those with ASD and ADHD, contribute distinct cognitive abilities to the workplace but may also share common challenges related to executive functions (EFs), such as working memory, cognitive flexibility, and organization and planning. These functions are critical for behavioral regulation, effective teamwork, and task execution in any workplace.

ADHD is associated with heightened creativity and divergent thinking, offering individuals with this condition the ability to generate novel ideas and solutions~\cite{hoogman2020creativity}. Similarly, individuals with ASD often exhibit unique cognitive processing styles, such as advanced spatial and object visualization~\cite{kozhevnikov2007cognitive}, which enhance their analytical skills and technical contributions~\cite{temple2022}. Recognizing these strengths, companies like SAP, Hewlett Packard Enterprise, and Microsoft have embraced neurodivergent hiring initiatives, leveraging these abilities to drive improvements in productivity, quality, and innovation~\cite{austin2017neurodiversity}. However, these benefits are often accompanied by challenges, particularly in areas such as social interaction, anxiety management, and perfectionism, which require thoughtful workplace accommodations.

From 2018 through 2022, the annual Stack Overflow Developer Survey~\cite{developerSurvey} collected demographic data on neurodiversity, showing a growing trend, with 4.27\% and 10.27\% (out of 71K+ respondents) reporting they had autism/ASD or a concentration and/or memory disorder (e.g., ADHD), respectively. This growth aligns with the increased awareness of Asperger’s syndrome being absorbed into ASD under the Diagnostic and Statistical Manual of Mental Disorders (DSM-5)~\cite{smith2020coming}, and the role of social media and online communities—especially after the COVID-19 pandemic—in aiding self-discovery and the pursuit of diagnoses~\cite{eagle2023you}.

Despite growing interest in the intersection of neurodiversity and software development, there remains a significant gap in the SE literature that require more empirical studies~\cite{marquez2024inclusion}. This interdisciplinary study, conducted by researchers from the Computer Science and Psychology fields, seeks to address this gap by examining the experiences of neurodivergent software developers, particularly those diagnosed with ASD and ADHD. While the manifestation of these two conditions can frequently co-occur, one individual may have one disorder without having the other~\cite{kern2015asd}. Using a Socio-technical grounded theory approach~\cite{hoda2024socio}, we conducted 25 interviews with neurodivergent developers to investigate the challenges they face, particularly those arising from the cognitive (e.g., EFs) and emotional characteristics inherent in their neurodevelopmental conditions, and how these struggles intersect with Agile practices. After results started to emerge during the analysis, we interviewed 5 neurotypical individuals to understand if they face similar problems. This research aims to identify the challenges faced by these neurodivergent developers and how adjustments in software teams and in the usage of Agile methodologies can better support their needs. Our findings have the potential to inform inclusive practices fostering a more supportive and productive environment for neurodivergent developers, enabling organizations to leverage these individuals' capabilities.


\section{Background and related work}

\subsection{ASD and ADHD comorbidity}

ASD and ADHD are neurodevelopmental disorders, which are conditions that begin during the developmental period~\cite{dsm-5}. They are marked by developmental deficits that impair personal, social, academic, or occupational functioning. These deficits range from specific issues with learning or EF to broader impairments in social skills or intelligence. However, the twice-exceptionality in some neurodivergent individuals, characterized by high achievement alongside disabilities, can confound the recognition of their cognitive difficulties, as their exceptional talents may mask areas of struggle~\cite{reis2014operational}.

ASD encompasses a broad range of characteristics and levels of support. Because it is a spectrum, people with ASD may present with a variety of characteristics, the combinations of which may vary from one individual to another. There are several traits such as difficulties in social interaction, communication and language, along with restricted interests, repetitive patterns of activities, and difficulty understanding concepts with ambiguous interpretations~\cite{dsm-5}.
Although ASD consists of a wide range of symptoms and levels of severity, there are many adults  who have independent and functional lives (i.e., high functioning autism)~\cite{howlin2017autism}. With growing understanding of ASD, more adults who never considered themselves autistic are now being diagnosed, leading to better recognition of lifelong challenges~\cite{happe2016demographic, lupindo2023late}.

ADHD is a disorder mostly related to impaired EF, i.e., ``executive dysfunctioning''~\cite{castellanos2006characterizing}. Executive functions are a broad term encompassing various cognitive processes, such as planning, cognitive flexibility (adapting to new information or rules), working memory (holding and manipulating information), attention, and inhibitory control (resisting impulses or distractions)~\cite{Goldstein2014}. In addition to these EF, ADHD impairment is also related to the so-called ``hot'' EFs that are associated with emotional and affective aspects of cognition, particularly those involving rewards and motivations~\cite{zelazo2002executive,castellanos2006characterizing}. For example, this distinction contributes to the common trait of delay aversion~\cite{sonuga1992hyperactivity} seen in many individuals with ADHD, characterized by a preference for immediate rewards over delayed ones of greater value. Finally, problems related to state regulation (SR) -- the ability to maintain an optimal physiological state of arousal that supports cognitive processes (i.e., EFs) -- have also been linked to ADHD~\cite{sergeant2005modeling}.


Scientific literature has evidence that 50 to 70\% of individuals with ASD have ADHD as a comorbidity~\cite{hours2022asd}. Since both disorders share a number of features and pathophysiological conditions -- especially related to EFs -- this may suggest a continuum and potential common origin~\cite{kern2015asd}. While most shared aspects among the two disorders are frequently seen around EF, there are also key ASD issues such as sensory processing difficulties and communication/social interaction aspects that have been reported in ADHD individuals~\cite{kern2015asd} as well. Although research on ASD and on ADHD have developed as distinct areas in Neuroscience, the similarity between them suggests these two research domains merge to investigate their co-occurrence further~\cite{vaidya2022comorbidity}.

\begin{figure*}[!ht]
    \centering
    \includegraphics[width=1\linewidth]{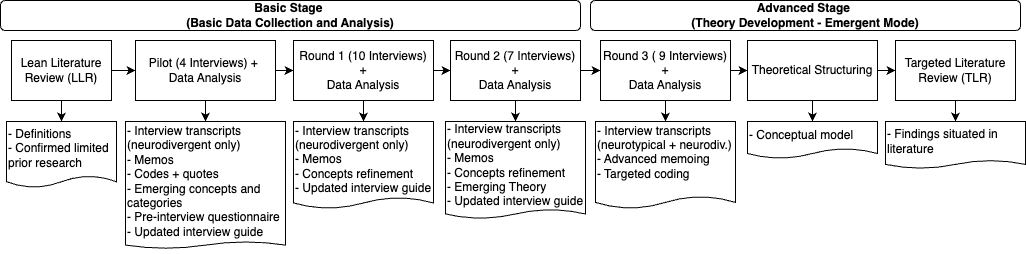}
    \caption{Socio-technical grounded theory method applied using basic and advanced stages. The main steps are the rectangles with the respective outputs below. }
    \label{fig:method}
\end{figure*}

\subsection{Research on Neurodiversity in Software Engineering} 
A recent multi-vocal literature review~\cite{marquez2024inclusion} highlights significant gaps in the inclusion of individuals with ASD in SE. Few studies identify facilitators like attention to detail and motivation, and these remain underexplored in real-world SE practices. Barriers  like communication difficulties, anxiety-inducing environments, and inadequate training are noted, but the effectiveness of interventions is rarely empirically tested. Proposals like immersive reality and divergent computational thinking remain theoretical, highlighting the need for empirical research focused on SE. The authors highlighted the need for further empirical studies more directed to that field. Overall, the reviews articles are secondary studies or education-focused articles with the exception of two studies~\cite{das2021towards, morris2015understanding} that have empirical evidence relevant to the context of Software Development.  Das et al~\cite{das2021towards} explored the work-from-home experiences of neurodivergent professionals during the COVID-19 pandemic, including but not limited to software developers. The study brings findings into the need for accessible workspaces, inclusive communication practices, and balancing productivity with well-being, but it is broad and has no focus on SE. Although Morris et al~\cite{morris2015understanding}  also bring general purpose accommodations that are applicable to many professional domains, they leave pertinent research questions for future work such as pairing neurotypicals with neurodivergent developers or adaptations that should be made to Scrum or any software development process in use.

A recent study~\cite{liebel2024challenges} highlights ADHD software engineers challenges like task organization, time management, and maintaining focus, which can lead to stress. However, strengths such as creativity, problem-solving, and systems thinking are also evident. The authors recommend adopting inclusive practices like flexible work modes, clear communication, and provide some hints around Agile methodologies to better support neurodivergent developers. The major drawback is the lack of neurotypical engineers among the interviewees. Finally, our pilot work that precedes the current study combined both ASD and ADHD software developers but is of limited and preliminary nature, reporting aspects on challenges such as remembering discussions from meetings, coping with sudden change, or dealing with impulsivity~\cite{gama2023understanding}.

\section{Method}
Socio-Technical Grounded Theory (STGT) is an iterative and incremental research method that combines traditional and modern techniques to conduct socio-technical research, aiming to develop novel, useful, parsimonious, and adaptable theories~\cite{hoda2021socio}.
We employed a full STGT study involving a multi-phase, iterative process of data collection and analysis, emphasizing both the technical and social aspects of software developers with ADHD and ASD. The method seemed appropriate since the intended data collection was qualitative and the topic predominantly has social and technical aspects interwoven~\cite{hoda2024socio}. The overall research question we intended to answer in this study is \textit{``How does the neurodivergent condition of ASD and ADHD individuals affect their work as software engineers in Agile software teams?''}.

\subsection{STGT Basic Stage} As illustrated in ~\autoref{fig:method}, we started with a Lean Literature Review (LLR) to define key terms and confirm the limited scope of existing research, allowing the research to remain grounded in participant data. Following this, four pilot 60-minute interviews were conducted with neurodivergent participants which were later transcribed and analyzed to generate initial transcripts, memos, and emerging concepts. As indicated by Hoda~\cite{hoda2024socio}, if the pilot was conducted with the same source cohort of the study, its data can be included in the study.  These initial results helped refine the interview guide to be used in subsequent rounds of interviews. The guide had many questions related to EFs and we decided to collect such data using an existing validated instrument that worked as a pre-interview questionnaire (PIQ)~\cite{hoda2024socio}. For that, we chose the IFERA-II~\cite{dias2021development} instrument, which focuses on difficulties in EFs, regulation and delay aversion. It is a validated scale that has advantages over the popular BRIEF instrument~\cite{roth2013assessment}, e.g., being more concise and also considering constructs (e.g., delay aversion) aligned with the most recent and accepted definition of EF~\cite{diamond2013executive}. 
In Rounds 1 and 2, neurodivergent participants were interviewed. The analysis in each interview round included open coding, constant comparison, memoing, leading to the identification and refinement of emerging concepts and categories as well as the  continuous refinement of the interview guide and the theoretical insights. As the process evolved, the analysis became more focused, and concepts were solidified, preparing for the advanced stage of theory development.

\subsection{STGT Advanced Stage} In the advanced stage, coding continued along with constant comparison, with the aim of identifying a core category that links all other categories and explains the central phenomenon under study, with targeted coding and deeper memoing being used to shape the emerging theory. During the theoretical sampling in this stage -- when the team follows the ongoing process of assessing the emerging codes,
concepts, (sub)categories, and hypotheses, and identifying new sources for
data collection~\cite{hoda2024socio} -- we decided to include neurotypical software engineers to fill theoretical gaps. Interviews in Round 3 included both neurotypical and neurodivergent participants. The IFERA-II PIQ was also used as a screening instrument with the neurotypical participants. After completing and analyzing the interviews, a conceptual model was structured based on the data, providing a grounded explanation on how the cognitive dysfunctions of neurodivergent engineers affect SE performance. Finally, a Targeted Literature Review (TLR) situated the emergent theory within the existing academic literature, aligning the model with established research findings and contributing novel perspectives to both technical and social dimensions of neurodivergent developers in software teams. This method ensures that the theory is deeply rooted in real-world experiences while it also integrates both the social and technical factors that shape these practices.

\subsection{Data Collection and Analysis}

Data collection was performed by two neurodivergent researchers (ADHD and ASD) who disclosed their condition when advertising the research and when contacting participants. This transparency fostered empathy together with a sense of rapport and mutual understanding, facilitating a more comfortable and open communication dynamic. While the pilot was a convenience sample through a referral chain, the invitation for the participation in the interviews was sent through a post shared on a social network (Twitter/X) and reposted by third parties, aiming at software developers from  \brazil. The form was filled out by 79 people stating they were interested. They were gradually contacted to schedule a video call interview. Interviews with neurodivergent individuals were scheduled as 60-minute slots, with some of them being slightly overtime, while neurotypical individuals (contacted through convenience sampling) had 30-minute slots. The subjects profiles are described in ~\autoref{table:participants}, mainly software engineers with varied experience. Concerning gender, the neurodivergent sample consisted of 18 men, 5 women and 2 non-binary individuals, while the neurotypical sample consisted of 3 men and 2 women. Work modes varied, with most of them working either as fully or partially (i.e., hybrid) remote.

\begin{table}[!h]
  \caption{Profile of study participants describing their position, age, diagnosis (ADHD, ASD or both) and time of diagnosis (years)}
   \label{table:participants}
\begin{tabular}{llllc}
\textbf{Part.} & \textbf{Position}             & \textbf{Age} & \textbf{Diagnosis }    & \textbf{Diag. time (y)}    \\
\hline
C1    & Sr. Software Eng.    & 34  & ADHD + ASD &     8        \\
C2    & Data architect       & 27  & ADHD + ASD &     7        \\
AD1    & Software Engineer    & 28  & ADHD       &     5        \\
C3    & Software Engineer    & 27  & ADHD + ASD &     2   \\
AD2    & Tech lead            & 29  & ADHD       & 1           \\
AS1    & Software Engineer    & 26  & ASD        & \textless 1 \\
AS2    & Software Developer   & 30  & ASD        & \textless 1 \\
AD3    & Software Architect   & 32  & ADHD       & 8           \\
AS3    & Tech lead            & 25  & ASD        & \textless 1 \\
C4   & Software Engineer    & 30  & ADHD + ASD & \textless 1 \\
AD4   & Sr Software Engineer & 46  & ADHD       & \textless 1 \\
AS4   & Project Manager      & 36  & ASD        & \textless 1 \\
AS5   & Sr Software Engineer & 30  & ASD        & \textless 1 \\
AD5   & Sr UX/UI Developer   & 52  & ADHD       & \textless 1 \\
C5   & Senior Developer     & 44  & ADHD + ASD & 1           \\
C6   & Software Engineer    & 34  & ADHD + ASD & 1           \\
C7   & Staff Engineer       & 37  & ADHD + ASD & 1           \\
AD6   & Software Engineer    & 34  & ADHD       & 1           \\
AD7   & Lead Engineer        &  39   & ADHD       & 1.5        \\
AD8   & Software Engineer    & 30  & ADHD       & 1.5         \\
AS6   & Software Engineer    & 34    & ASD        & 6           \\
AS7   & Software Engineer    & 24  & ASD        & \textless 1 \\
C8   & Test Automation Eng. & 27  & ADHD + ASD & 15            \\
AD9   & Software Engineer    & 32  & ADHD       & 15            \\
AD10   & Software Engineer    & 26  & ADHD       &  2           \\
N1   & Software Engineer    & 23  & Neurotypical        & -         \\
N2   & Software Engineer    & 23  & Neurotypical        & -         \\
N3   & Software Engineer    & 23  & Neurotypical        & -         \\
N4   & Sr Software Eng. & 42  & Neurotypical        & -         \\
N5   & Software Engineer    & 27 & Neurotypical        & -        
\end{tabular}
\end{table}

 We transcribed interview recordings with an automated tool and later verified the results to adjust minor errors from that process. As part of an iterative approach, the analyses were made in batches~\cite{hoda2024socio}. The raw data was open coded with hashtags, as stated by Hoda, ``to conceptualise and represent a key idea in a parsimonious way''~\cite{hoda2024socio}. Also following her recommendations, we attempted to apply the zoom out-zoom in approach by closely examining the individual experiences and interactions of neurodivergent software engineers with their work environments and practices (zoom-in), while zooming out to also connect them to broader social (mainly under Psychology lenses) and organizational practices. Through constant comparison of codes, this led to concepts, which were clustered together to create categories, as exemplified in ~\autoref{table:coding}. Supported by a shared online whiteboard, coding was mostly done by one researcher along with frequent video calls among the other researchers for validation and to converge opinions in the clustering that lead to concepts and categories.

\begin{table}[]
\caption{Coding evidence of three concepts that are part of the Cognitive and Emotional Dysfunctions category}
\label{table:coding}
\begin{tabular}{p{6.5cm}p{1.55cm}}
\textbf{Open Code + Raw Data}  & \textbf{Concept} \\
\hline
\hline
\#\textbf{anxietyWhenBreakingDownTasks} \\ ``The task is big. Until I manage to break it into smaller tasks, there’s a whole process [..]
. That generates a lot of anxiety.''- AS1  & Organization and  planning  \\
\#\textbf{frustrationWithTaskOrganizationTools} \\ ``I basically can't handle organization tools very well, it's always a love and hate relationship.'' - C4                                                                          &      
\\
\hline

\#\textbf{taskInitiationThroughExternalPressure} \\ ``What makes me start a task is being asked about it. I think that’s it.
''- AS1                                                           & Task initiation                                               \\

\#\textbf{procrastinationOnUnfamiliarTasks} \\ ``I procrastinate mainly if it’s a task I’m not very sure how to do, something new [..].
'' - C8\        &                                                               \\
\hline
\#\textbf{disruptionOfFlowCausingFrustration}\\ ``Sometimes I catch the flow and just want to follow that line of reasoning. If something breaks that line of reasoning, I get very upset. 
'' - C6 & Emotional dysregulation                                      \\
\#\textbf{rejectionSensitivityFromCodeReviews}   \\ ``You already think that you're being rejected, that you're not a good professional, that you don't know.'' - C4                                                                  &                                                 
\end{tabular}
\end{table}

The Inventory of Difficulties in Executive Functions, Regulation and Delay Aversion for Adults (IFERA-II) instrument allowed the comparison between groups (ASD, ADHD, ASD+ADHD and neurotypical) using descriptive statistics, since the sample was small under a quantitative perspective. The average scores for each construct are depicted in \autoref{fig:iferaii}. The data collection instruments and a sample code book with results from the data analysis are available online\footnote{\url{https://figshare.com/s/4fac07f4f00b348346f6}}.

\subsection{Ethics}
The project was approved for by the lead author's university ethics committee, being in accordance with national laws regulating research involving human subjects. Each subject would participate only after filling out the informed consent form. Participants were informed about potential emotional triggers during the interview and the possibility of abandoning it at any moment. Data was allowed to be shared only as summarized information (i.e., not at the individual level).

\section{Theory}
\begin{figure}
    \centering
    \includegraphics[width=1\linewidth]{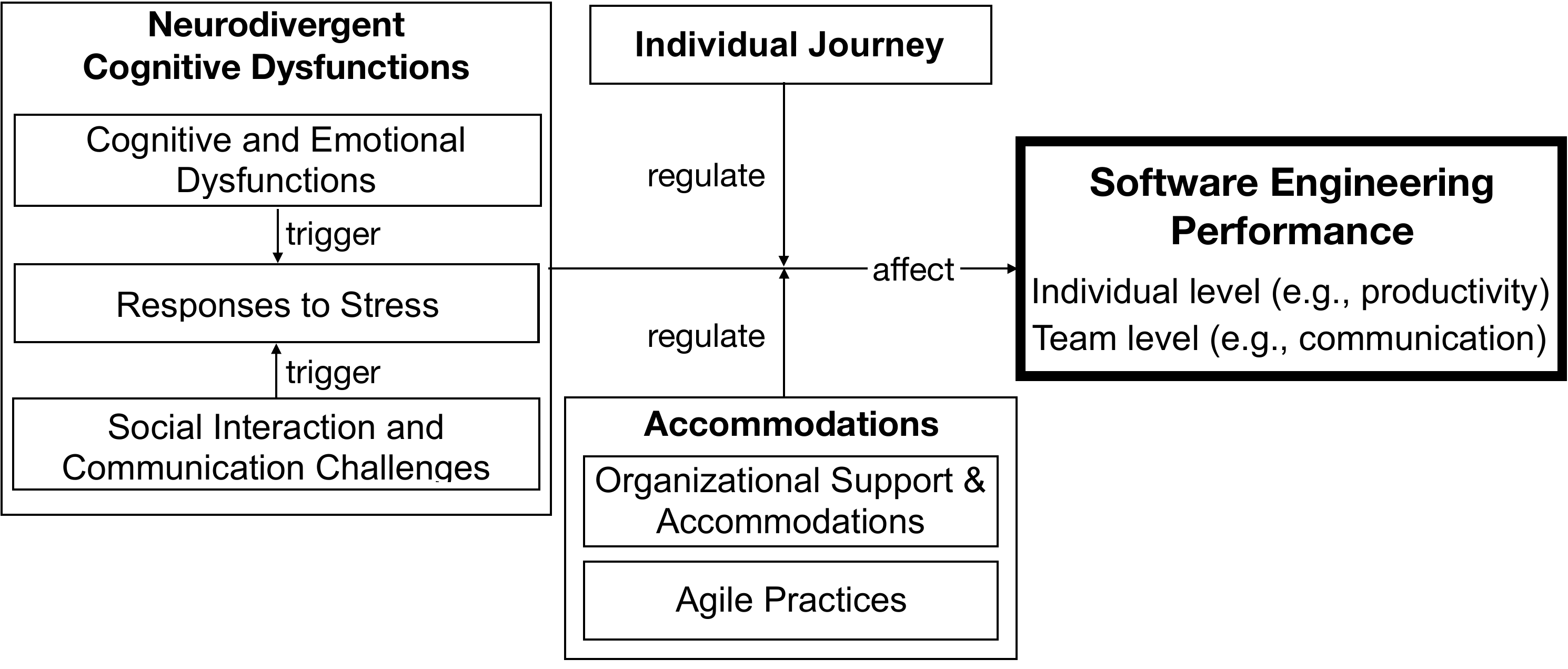}
    \caption{Theory on the Effect of Neurodivergent Cognitive Dysfunctions in Software Engineering Performance}
    \label{fig:theory}
\end{figure}

As the result of our theory structuring, we reached a conceptual model (\autoref{fig:theory}) that represents our theory around a core category that represents the Software Engineering Performance -- both on individual and team level -- of neurodivergent individuals, being affected by various cognitive dysfunctions. Specific dysfunctions of cognitive (e.g., EFs) and emotional nature can trigger responses to stress, such as anxiety and mental exhaustion (e.g., burnout). This stress can also be triggered by social interactions and communication challenges that are part of those neurodivergent cognitive dysfunctions. The Individual Journey of this individual, which encompasses diagnosis, therapy, medication and self-awareness, as well as various Accommodations (organizational support and Agile practices) can regulate how these dysfunctions affect their SE performance. The next subsections describe the categories that comprise that model and the respective concepts that compose them. Note that the effect on SE performance is described together with the Neurodivergent Cognitive Dysfunctions, as our interviewees would report how a specific dysfunction or challenge would affect their performance.

\subsection{Cognitive and Emotional Dysfunctions}
\paragraph{Organization and planning}
Although Agile practices value responding to change over following a plan~\cite{AgileManifesto}, planning is still relevant. For individuals with ASD or ADHD, it is both necessary and something they often struggle with. \textit{``The execution tasks are the easiest, and the planning tasks are the hardest.''} - AD5. Supporting task clarification can make life much easier for individuals with ASD. \textit{``I have trouble doing this alone. Even if it’s already divided, sometimes I feel the need to talk to the person who’s reporting the task and have a conversation to get that understanding.''} - AS7. Clear instructions make a big difference. \textit{``If I have clearly the guidelines of what I have to do, it's much easier for me to solve a problem.''} - AS2. Many interviewees reported difficulties in breaking down tasks. \textit{``When there's something very big, if I don't break it into smaller things, I easily get lost.''} - C6. 

\paragraph{Task initiation challenges}
Several interviewees report their struggles in initiating tasks. \textit{``Tasks that you don't have much desire for, at least for me, I don't know how other people are, but with a boring task the problem is that the more you imagine how much work it will take the less desire (to do it) you have.''} - AD2. 
Adequate levels of pressure set by deadlines or an accountability partner can help overcome procrastination. \textit{``What makes me start a task is being asked about it... I think that’s it. Deadlines and someone asking about it.''} - AS1.

\paragraph{Perfectionism and overengineering}
In Agile development, delivering functionality quickly lessens the focus on aspects such as design, good programming practices, or test coverage, leading to technical debt~\cite{behutiye2017analyzing}. This is relevant for our population, as many subjects struggle with perfectionism, a commonly endorsed cognitive distortion in ADHD~\cite{strohmeier2016assessment}, leading to overengineering, in the opposite way of Agile. \textit{``He said, no, you're too perfectionist. It has to be delivered. It's not that it has to be bad.''} - C5. As developers become aware of their tendency for overengineering, they devise strategies to mitigate this risk. \textit{``Something I learned is how to work around it. If I see it's an urgent task, I don't do that overengineering anymore. I see it needs improvement, and I open another task for later when I have more time.''} - AS6.

\paragraph{Irregular work performance}
Several of our interviewees mentioned they would prefer flexible working hours. Sleep disorders are common in both ASD and ADHD and may be related daily impaired functioning~\cite{lugo2020sleep}. \textit{``I feel like I function better at night. In the morning, I’m really sleepy, really slow.''} - C8. While this preference is in line with their struggles concerning focus and attention, some of our interviewees noted how this has a negative health impact. \textit{``My psychiatrist said this is very wrong. I always end up stretching my work into the night.''} - AD10. 

\paragraph{Emotional dysregulation and intrusive thoughts}
The often-mentioned ability to focus from our respondents has a downside: they can feel frustrated when that focus state is interrupted. \textit{``Sometimes I catch the flow and just want to follow that line of reasoning. If something breaks that line of reasoning, I get very upset. Like, I get internally disorganized.'' - C6}. This can generate tensions within the Agile team. Feeling accepted within the work context is extremely important for professionals in general, and our respondents are no exception. \textit{``This rejection issue. Today I can understand and suffer less. Because I wanted to be accepted, I just wanted to be accepted.''} - C5. Rejection sensitivity is particularly prevalent among our respondents. Ordinary development activities (e.g., receiving a harsher code review) can trigger such feelings.

\paragraph{Low self-esteem}
As seen in literature~\cite{newark2016self}, some of our neurodivergent interviewees seem to have a negative bias against themselves, even when external evidence suggests otherwise. \textit{``It’s a very recurring issue, that either what I did isn’t good enough, or even when I get good results, I think that one day they’ll find out I’m a fraud.''} - AD10.

\paragraph{Difficulties with context switching} Several interviewees noted how they struggle with context switches. \textit{``if you try hard to grab my attention, eventually I'll look. The real problem is getting back to context. It's not just about losing focus, but about switching back to the original context.''} - AD2.

\paragraph{Cognitive flexibility} Responding to change is a key Agile value. Neurodivergent individuals often face change \textit{``not without a bit of anxiety, depending on the level of change''} - C4, and find it stressful: \textit{``When I got the news about the project change, I was kind of paralyzed.''} - C8. Embracing change can bring discomfort \textit{``No one's going to convince me, basically. I'll do it because I have to get the work done, but I'd be unhappy doing it.''} - AS7. In spite of some rigidity in the work process \textit{``Some things have always to be the same way''} - AS5, \textit{``an autistic person accepts the change if it's grounded in a solid foundation''} - AS2. Having time and resources helps adapting to change \textit{``I said, `no, it's not all good. I'm getting anxious.' But it's because it's a change of plans [..] I think it's about giving people time to adapt, you know? To have brief documentation.''} - C4. Developers end up adopting strategies to cope with change: \textit{``Getting stuck on something and spending hours trying and failing to find the solution. But I'm becoming more self-aware and finding ways to work around it.''}- AS7. 

\paragraph{Focus and attention}
Several participants report having ``hyperfocus'', a mental state of sustained, deep concentration typically associated with ADHD and ASD~\cite{ashinoff2021hyperfocus,grotewiel2022experiences}. 
For some, hyperfocus can be perceived as an advantage: \textit{``It was more of a blessing than a difficulty, because I could explore the possibility of working on extremely interesting problems, which were in full harmony with my hyperfocus.''} - AS2. 
However, hyperfocus can also lead to one neglecting basic needs like eating \textit{``I would enter hyperfocus in the morning and finish work at night without having lunch''} - C2, sleeping \textit{``There were several times I was working, like at night, when suddenly I say `Oh my God, it's already daylight, I didn't sleep!'''} - AD3, or even affecting personal hygiene \textit{``I even forget the basic physiological needs, I sometimes forget to shower.''} - AS2. This can have severe health consequences \textit{``I developed diabetes not just because of the family history, but also because I sometimes forget to eat due to hyperfocus.''} - C4. 
Hyperfocus can also lead to scope creep. \textit{``The priorities are these, and hyperfocus comes in, and I forget about the world and end up with something much bigger.''} -AS5. 

Another side of this coin is the lack of concentration on other issues, as subjects lose focus on topics they find uninteresting: \textit{``If someone starts talking about something that doesn't interest me, I lose focus. If I start focusing on something else, sometimes I go back to something I'm interested in, and I'm not even paying attention anymore.''} - AS3. This makes long meetings particularly challenging: \textit{``Sometimes I'm in a meeting that doesn't seem all that useful to me, but I'm there, so I start listening and get lost. By the end of the meeting, I sometimes leave without knowing what was discussed.''} - AD7. 

Sensory overload can negatively impact focus: \textit{``I prefer environments with less light. Too much light disrupts my concentration. Repetive sounds also disturb me''} - AS1. They can also \textit{``interfere with emotional regulation''} - C6,  and indirectly damage social interactions \textit{``I don't think I get overwhelmed by social interaction, but by the stimulation that social interaction brings. Noise, lights, etc.''} - AD10.

\subsection{Social Interaction and Communication Challenges}
Various challenges of our interviewees related to social interaction and communication with other individuals
\paragraph{Social anxiety in the workplace}
Interviewees describe how they avoid social interaction at the workplace, and how this interaction makes them anxious:
\textit{``Before the meeting with my leader, where I had to say what I’d done during the week, there was always that anxiety, like, what excuses will I come up with now, what will I make up.''} - AS3.
\textit{``The daily meeting often causes anxiety, because if the day before wasn't very productive, if you procrastinated, you feel the weight of having to show your progress.''} - AD7  

For some interviewees, this relates to social activities that companies do as a part of the work:
\textit{``I have no problem recording videos; I'm used to pitching startups. What I can't do is these little games, dancing.''} - AS4. 

\paragraph{Body language and social cues}
Several interviewees with ASD have difficulties reading body language or other social cues during communication.
Among other things, this causes extra mental load.
\textit{``I think there was already prejudice because I received a lot of feedback about communication. For example, I often got comments like, 'Oh no, you need to read the room, you need to understand.'''} - C7.

\paragraph{Preference for textual communication}
For various reasons, such as difficulty reading body language and social cues, or sensory discomfort, interviewees mentioned that they have a preference for textual communication over in-person communication:
\textit{``And here it's calm because I'm at home and communication is only by text, I message and I can easily ignore.''} - AD3.
\textit{``When there's a group meeting, I end up using the chat more, because I'm afraid of interrupting people and talking too much.''} - C4.

However, one interviewee with ADHD specifically mentioned that they preferred verbal communication for clarifications:
\textit{``I reach out to the person, 'Do you have five minutes so I can ask something about the PR?' Then I try to explain it verbally, because I think it's easier than sending the wrong impression through writing.''} - AD8.  

\paragraph{Avoidance of Direct Contact}
Similar to avoiding in-person communication, four interviewees mention that they avoid direct contact with others, in particular due to difficulties in maintaining eye contact.
\textit{``In remote settings, I feel a bit more comfortable looking at the camera than looking at a person.''} - AS3. 
\textit{``You're talking to someone and, most of the time, looking at the desk or something else, rather than at their face. So, you're making less effort to look into people's eyes, which feels more natural.''} - C4  


\paragraph{Communication struggles}
As a final concept, we observe various struggles related to communication among our interviewees.
These struggles can take various forms, such as being too direct, becoming unresponsive, interrupting others, asking for help, or difficulties controlling the tone in a conversation:
\textit{``I have a lot of trouble asking for help, not just in development, but in everything... I feel like I just can’t ask for help.''} - AS1.
\textit{``But there are moments when I realize that I was too direct.''} - AS6.  
\textit{``I don't know how to identify, but there are some periods when I become quite, I don't respond to anyone.''} - AS5.
Interviewee AD8 is concerned with the tone of messages, and asks for external support to check it: \textit{``I ask my wife to read it, and sometimes I put it into ChatGPT to tell me the tone of the message''}.
Two interviewees went for training/mentoring to improve communication skills. AD3 gave up the training for not seeing significant advances  while C7 succeeded in improving his communication skills. 

\subsection{Responses to Stress}
As a result of cognitive and emotional dysfunctions, as well as social and communication challenges, our interviewees experience various stress responses. 

\paragraph{Mental exhaustion}
Interviewees experienced mental exhaustion after intense work period, especially burnout.
\textit{``And we can't handle a very large and tight flow without questioning all this. It comes with burnout and our burnout isn't their burnout.''} - C5.  
\textit{``I frequently have shutdowns when there are too many meetings. One day with back-to-back meetings. It seems fine, but by the end, I can barely speak. I try to remember a word, but I can't.''} - C6.  
\textit{``He realized that switching me from project to project in short periods kept me very productive. So he kept making these switches, and I ended up with hyperstimulating work that led to burnout.''} - AD10.  

\paragraph{Anxiety}
Many interviewees experienced anxiety for various reasons during their work.
Typical reasons include upcoming meetings (where they might have to speak), being evaluated, or simply receiving notifications.
\textit{``Not knowing what it’s going to be, what I’m going to say, or where the conversation is headed makes me a little anxious.''} - AD6.  
\textit{``If there’s some important event scheduled in the near future, I get into that mode where I start working, but in my head, it’s like, 'I need to do this.' And then my focus goes elsewhere.''} - C8.  
\textit{``They get in my way, to be honest. I’m here coding, something pops up, and I get anxious. I can’t forget. I’ve already gotten feedback, so I can’t forget.''} - AD8.  

\subsection{Agile Practices}

\paragraph{Pair programming}
Pair programming was considered in a broader sense of pairing up with someone and not necessarily having the driver and navigator roles enforced by XP. This seems to be a key technique to support task initiation, a major struggle for ADHD individuals.  
It was even seen as useful for neurotypical (N1) to start tasks: \textit{``When I’m working in pairs, I can work better. I feel things are more productive.''} - N1.
However, it can be seen with both benefits and drawbacks by autistics, who are affected by the social aspect of it, draining their ``social battery'', preferring to pair up with someone they know. 
\textit{``those three hours of social work will really wear me out, and I can’t do anything afterward.''} - C7.
\textit{``The issue of being observed. I would have to feel more comfortable with someone I know.''} -AS5.
Another approach by some people to start tasks was creating external accountability over an open online call \textit{ ``An open call on (Google) Meet. I enter the call, there are four people. So, when someone unmutes and says, 'Hey, can someone help me with this thing here?' and talk about some problem they had. It’s like, just unmute and talk.''  } - AS3.
\textit{``Not only about clarifying technical content, but I've also done this with some friends, like calling, staying in Gather Town, Meet, and staying in your own corner, quietly working, sometimes opening a microphone to comment on something, ask a question.''} - C4.  

\paragraph{Test-driven development}
TDD  naturally emerged in some interviews from both ASD and ADHD developers and it was confirmed in some interviews when asked. It was seen as a support for task initiation and for organization and planning . 
\textit{``TDD, in some cases, helped me start tasks. I don’t use it rigorously 100\% of the time, but in many cases, I managed to.'' }- C7.
\textit{``A problem that's hard for me to see a solution. It (TDD) helps you start thinking about the design''} -
AD4.

It also gives notion on the dimension of a task, if there still something to be done or if it is concluded:
\textit{``if your code passes all  tests, you’re done. So, it gives you a lot of clarity about how much is left and when to stop as well''}- AD9.

\paragraph{Tasks estimates and breakdown}
Participants mention the importance of overestimation of activity effort to create buffers and reduce stress: \textit{``So if I really mess up and say 'damn, it was a boring task, I ended up taking much longer,' I activate the time of the others that ended up giving extra fat to it.''} - AD2. 
\textit{``Everything I do, I always give a much longer deadline than I think I actually need.'' }- AD7.

Overall, the idea of breaking down larger tasks into smaller ones is key to have a notion of clear and achievable goals. In the absence of processes for activity planning, individuals  with ASD struggle more: 
\textit{``We have a lot of refinement, daily planning, and all that, and I see that in the companies where these flows and tasks aren't well-defined, I struggle more.''} - AS6. The definition of clear steps to develop the activity helps to reduce the typical ADHD procrastination:    \textit{``If it’s a slightly bigger task that requires step-by-step actions, and I don’t have that, then there will be procrastination.''} - C8.

\paragraph{Daily meetings and other practices}
Daily stand-ups provide a regular touchpoint for team members to align on tasks and progress, and can also act as an accountability mechanism, encouraging individuals to start tasks they might otherwise procrastinate on: \textit{``it's a great point of dailies, because I'm a person who has many social difficulties, so they're just a way for the team to come together.''} - C4 
\textit{``The fact that there’s a daily meeting forces you to start something, even when you don’t want to.''} - AD7.

Extended or poorly managed daily meetings can lead to frustration and wasted time, and repetitive updates without actionable outcomes can lead to disengagement: \textit{``It's a lot of meeting time, if it were 15 minutes, and knowing that it will be 15 minutes.''} - AD2.

Sprint plannings help in organizing tasks, setting clear goals, and providing a sense of accomplishment upon completion: \textit{``I like to have a (sprint) planning, where we really organize what we're going to do. I like the (sprint) review, when we deliver, it means it's a relief. I'm delivering because I'm completing what I planned to do.’’} - AD5.

Retrospective Meetings provide a platform for feedback, allowing team members to voice concerns and seek clarity on processes, which is something valued by ASD individuals, but they may also be seen as something that do not lead to actionable changes or meaningful outcomes: \textit{``Retrospective meetings give you a voice to say something you don’t like, something that isn’t very clear.’’} - C7. \textit{``There are types of ceremonies that, for me, don't work. Retrospectives, for me, I don't see a purpose, I don't see them adding much value, or having an impact on what we do.’’} - AS6.

Code reviews help identify issues that might be overlooked, particularly beneficial for individuals with ADHD who may miss details: \textit{``Code Reviews also help a lot, because for someone with ADHD, you let a lot of things slip by, so it ends up being really helpful.''} - C7. However, some individuals find code reviews stressful or are unsure whether they are beneficial in the long run: \textit{``It (code reviews) bothers me a bit, but I don't know if it helps or hinders in the long run.''} - AS5.

\subsection{Organizational Support and Accommodation}

\paragraph{Structural/Physical adaptation}
Providing isolated or quieter workspaces helps reduce distractions, enabling better concentration and focus through controlled environments. Allowing control over environmental factors such as lighting and temperature helps accommodate sensory sensitivities.
\textit{``Many places are noisy. I don’t like working right next to people; I prefer some isolation, or at least being a bit further away.''} - AS6. 
`\textit{`I realized that this was sensory overload. I also try to isolate noise, modulate how much exposure I have to light. At home, it's kind of dark. I'm fine in the dark without anything on.’’} - AS4. 
However, not all organizations can provide adequate sensory-friendly spaces, leading to unmet needs. 
\textit{``People needed more focus, they stayed, at certain points, more isolated. I also used these rooms a lot, only that there weren't enough for everyone who needed focus.'' }- C4.
    
While isolation can reduce distractions, it may hinder collaboration and social interaction: 
\textit{``If I’m at work, I want to collaborate, talk to others, and discuss things. So if I’m in a very closed environment free of distractions, I think it defeats the purpose of being in the office. I could just be doing that at home.''} - AS1. The physical act of commuting can add stress and sensory overload, exacerbating cognitive challenges.
\textit{``Commuting is a big overload for me, whether it’s traffic, crowded public transport, or just the time it takes to get from home to work.''} - AS6.

\paragraph{Team/Process Adaptations}
Leaders who recognize and adapt to neurodivergent needs can significantly enhance employee well-being and productivity. 
\textit{``My last manager told me, 'On bad days, let me know, and I’ll try to rearrange the team. If you're handling something urgent, we’ll reassign it.'''} - AD8.  
Limiting the number of meetings allows individuals to dedicate more time to focused, individual work. \textit{``In solo work, I end up doing more at the end of the day, like early afternoon, almost late afternoon, when there aren’t as many meetings, that’s when I manage to focus on my individual work.''} - AS3. Breaking down tasks into smaller, manageable units aids in organization and reduces overwhelm: \textit{``I like to have a (sprint) planning, where we really organize what we're going to do. I like the (sprint) review, when we deliver, it means it's a relief.'' }- AD5.

Variability in how team/process adaptations are applied can lead to unequal support among employees. \textit{``I've worked with someone who didn't have ADHD, who was super organized and everything. We worked well together and complemented each other. But I've also had someone who wasn't even diagnosed yet, and I had to push the other person, which demanded energy from me because I had to deal with my energy, but I had to push the other. It was more complicated.''}- AD5. 
Inclusion and accommodation policies need to be institutional. Informal support systems may not be sustainable or sufficient to meet all neurodivergent needs. \textit{``There's a certain need from the company about this, and I scheduled some talks with HR to discuss this because she just joined and had no experience about neurodiversity.'' }- AS4.

\paragraph{Individual Adaptations/Strategies}
Interviewees reported personalized coping mechanisms that are working for them. Externalizing some cognitive processes by leveraging tools like calendars, alarms, and note-taking aids helps manage time and tasks effectively. `\textit{`After I put all my calendars on my watch and started scheduling everything in Google Calendar, I no longer have that problem.''} - AD10. As explored in other mapped concepts, dividing tasks into smaller, achievable steps enhances focus and reduces overwhelm. \textit{``The smaller the task, the easier it is for me to do it in that attention span. So that helps me a lot.''} - AS1. 

As focus and attention strategies, incorporating regular breaks prevents burnout and maintains productivity. \textit{``When feeling overwhelmed, I try to breathe, take a few minutes off. As most of my work is remote, I can take short breaks.''} - AS6.

However, constantly managing personal strategies can be exhausting and may generate cognitive overload from actual work tasks. The inconsistent use of personal strategies can lead to missed tasks and increased stress too. \textit{``I think the most tiring part of ADHD is that you have to have various methods to absorb content. So you have to draw, you have to write, [..] you have several post-its of things [..] I tried to make a Planner also on the tablet and it didn't work, because I need something in my view to register things.''} - C4. 

\paragraph{Stigma and Ableism}
This concept represent challenges and negative experiences for neurodivergent individuals in the workplace. Recognizing these issues through organizational awareness is a path to overcome them.
Neurodivergent individuals often receive unfair performance evaluations and face pressure to conform to neurotypical standards. \textit{``They gave me a poor performance rating. Below the expected performance. I was having a crisis, I was under treatment. They tried everything to make me resign.''} - C5.
\textit{``"When it comes to a promotion, I feel like it will be a bit harder for me than for a neurotypical person.'' }- C7.
Many people still still see stereotypes that question the capabilities and social skills of the neurodivergent.
\textit{``I don’t feel very comfortable sometimes with people treating me like, not exactly like a child, but like I’m incapable.''} - C8.
\textit{``I think when we open up about this, people form very shallow opinions.''} - AS4.

\paragraph{Organizational Awareness}
The awareness about neurodiversity should be at the institutional level so the organization can structure it: \textit{``We have what we call an affinity group, for people with disabilities and neurodiverse people.''} - C7. 

Organizations that embrace flexibility can better accommodate the varying needs of neurodivergent employees: \textit{``Actually, there were accommodations because, since the schedule is flexible, it's understood that we can  adapt the work environment in the best way for the people.''} - C4.

Although the personal experiences of AD8's leadership brought empathy toward ADHD, if such awareness from leadership is in the whole organization, a more empathetic environment can be fostered: \textit{``My leader knew I had ADHD before I even told him. Because he had already noticed traits in me that he saw in his daughter.''} - AD8. 

\subsection{Individual Journey}

\paragraph{Neurodivergence Diagnosis and Comorbities}
The beginning of the individual journey starts with the diagnosis, which were reported from different paths such as recognizing themselves in YouTube videos, seeing symptoms in their recently diagnosed children, recommendations during sessions with mental health professionals. \textit{``First, YouTube diagnosed me by starting to suggest videos about autism, exceptional abilities, autism in adults, autism in women. [...] I identified with a large number of the signs there. Then, it was confirmed with the autism and ADHD diagnoses.''} - C6. There were many cases of comorbidities, not only the co-occurrence of ASD and ADHD, but also other psychological disorders: \textit{``General anxiety disorder and also diagnosed with obsessive-compulsive disorder.''} - C4. A twice-exceptionaly (i.e., neurodivergent with intellectual giftedness) diagnosis was reported by 7 subjects. 

\paragraph{Medication and Therapy}
While ASD interviewees took medication only for co-morbidities (anxiety, depression), ADHD medication has significant impact in helping to regulate EFs. Many reported taking Lisdexamfetamine or Methylphenidate, but the the former was frequently cited with side effects: \textit{``I had a mood rebound, I got very depressed.''}- AD9. In the case of therapy, more than half (13) of the neurodivergent participants mentioned Cognitive Behavioral Therapy (CBT). One ASD subject mentioned it did not work for them:\textit{ ''I do Cognitive Behavioral Therapy, but CBT doesn't work very well.''} -AS5. Approaches such as schema and occupational therapy were mentioned to a lesser extent.

\paragraph{Self-awareness}
After having the diagnosis, in combination with meditation or therapy, most interviewees started a path to understand better their condition, which helps them with self-acceptance.
\textit{``Understanding the diagnosis, understanding the traits, discovering myself, seeing my strengths and weaknesses helped a lot in my professional career.'' }- AS6.

\paragraph{Family and relationships}
Mentions to family had connections with the awareness of being neurodivergent, somehow helping to explain where it came from, while relationships were mostly about the ASD difficulties of engaging in a relationship.
 \textit{``Because my mom has it, my brother has it, and he has it quite intensely. The whole family on my mom's side has it, like five or six diagnosed people.''} - C8.

\section{Discussion}
In the following, we discuss our findings and provide concrete recommendations for SE practice.

\subsection{Neurodivergent x Neurotypical}
\begin{figure}
    \centering
    \includegraphics[width=1\linewidth]{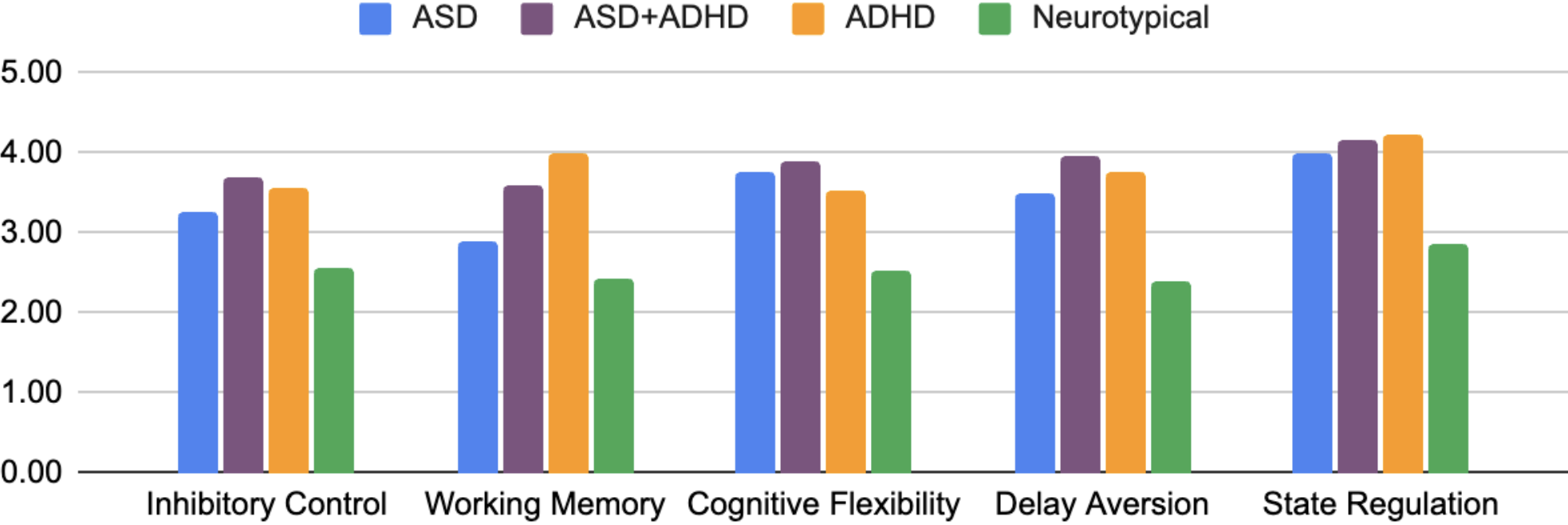}
    \caption{Average of participants answers to the instrument used after the pilot to ask about EFs, regulation and delay aversion inventory. ASD (n=7), ASD+ADHD (n=5), ADHD (n=9) and Neurotypical (n=5)}
    \label{fig:iferaii}
\end{figure}

In general, the interviews confirmed that neurotypical software engineers do not have the same dysfunctions as theirneurodivergent counterparts. Aspects related to EFs were no major concerns for them. While some complained about boring tasks, this was not to the level of engineers with ADHD. One neurotypical respondent had more evidence of delay aversion and emotional dysregulation on negative feedback, which affected the IFERA-II scores due to the small sample size.  Overall, the comparison of EF (Inhibitory Control - IC, Working Memory - WM, and Cognitive Flexibility - CF), state regulation (SR), and delay aversion (DA) across neurotypical individuals (ASD, ADHD, and ASD+ADHD) reveals distinct patterns of cognitive and emotional challenges (see \autoref{fig:iferaii}). Neurotypical individuals report the lowest levels of difficulty across all constructs. When observing the neurodivergent, SR had a similar score for all three groups. ADHD reported highest difficulties in WM. In general, ADHD scores were higher than ASD in IC, WM and DA. These scores align with the known impulsivity and difficulties in maintaining attention and controlling emotional states in  ADHD~\cite{zelazo2002executive,castellanos2006characterizing,sonuga1992hyperactivity}. CF was the only EF where ASD scored higher than ADHD, consistent with the characteristic need for routine and difficulty in adapting to change often associated with ASD~\cite{geurts2009paradox}. In contrast, the ASD+ADHD group scores high in all constructs. It reinforces the importance of increasing studies on the co-occurrence of those disorders, as already pointed out~\cite{vaidya2022comorbidity,kern2015asd}. The results suggest that neurodivergent software engineers face significant cognitive and emotional regulation challenges that are likely to impact their performance in structured environments such as work.

\subsection{Implications for Practice}
Our findings have implications on how to create inclusive and supportive environments for neurodivergent individuals in SE, facilitating their integration and productivity within Agile teams. Specifically, we make the following recommendations.

To \textbf{make sprint planning more inclusive}, ensure that task breakdown is clear and explicit for neurodivergent individuals. This reduces their cognitive load and can improve task comprehension and completion rate. As a specific practice, we find that TDD has the potential to help as a strategy to break down a task into tangible and smaller chunks.

Among the tasks planned for a sprint, \textbf{explicitly identify and highlight tedious, repetitive tasks}, as neurodivergent individuals might have a hard time completing them. For these tasks, recommend pairing a neurodivergent engineer with someone else to maintain engagement and productivity.

\textbf{Provide communication training and mentoring} for teams that contain both neurotypical and neurodivergent team members. We believe it is essential to not only provide this training to neurodivergent individuals, but also improve how neurotypical team members communicate with their neurodivergent colleagues.

\textbf{Keep teams as stable as possible}, so that team members with ASD work with familiar colleagues. 
An important observation we made is that for many of our interviewees, the familiarity to other people is a key aspect.
That is, while they typically struggle in social situations or in communication, this is not necessarily the case with people they know.
This highlights the importance of stable routines and few changes to a team.
The Agile team can, over time, become such a place of stability where neurodivergent individuals can also perform well in terms of communication and social interaction.
We believe this is an important first step towards creating ``safe spaces'' for neurodivergent individuals in companies.
As a result, this recommendation also entails spending substantial effort on onboarding for new team members.


\textbf{Allow for written communication}, even in co-located teams. This accommodates neurodivergent team members who may have difficulties with oral communication, even if only temporarily.




Finally, \textbf{shorten the duration of pair programming sessions, daily stand-up meetings, and retrospectives}, to reduce the chance of cognitive fatigue. In the same direction, ensure that planned meeting times are kept. 

\subsection{Findings Compared to Literature}

STGT suggests a targeted literature review to compare the findings of the study highlighting differences, similarities, and contributions to the field's progress~\cite{hoda2024socio}. As in the SE field this is a very novel topic, we focused on white papers about neurodiversity in the workplace. The selected white papers~\cite{wp_bankofamerica2023,wp_deloitte2022,wp_everymind2021,wp_ucu2022,wp_mcdowall2023neurodiversity,wp_thompson2024, wp_adhdfoundation} have many common aspects among them, such as quiet workspaces with sensory adjustments, flexible working arrangements, personalized career development, flexible workspaces and remote options, neurodiversity training for leaders and manager. Accomodations and adaptations such as coaching and support for planning and organization, clear communication and structured tasks had more relation to our findings around agile practices.

The work of Morris et al.~\cite{morris2015understanding} pointed out future directions that were not taken into practice and that in a certain way overlap with our findings. They suggested studies on (1) evaluating the benefit of pairing up with neurotypical individuals under a perspective of code quality; (2) making changes to scrum or other processes to support neurodivergent participation; and (3) investigating tools to interpret nuance within messages. Concerning (1) we observed the importance of that practice playing a role on task initiation and sustained attention. In the case of (2), we point out a tangible list of practices to be adapted based on empirical evidence, and for (3) we recommended improving communication as a soft skill, but also providing textual communication tools. Additionally, our sample showed anecdotal evidence of ChatGPT being used to check the tone of written communication. 

Related to the work of Liebel et al.~\cite{liebel2024challenges} about individuals with ADHD in SE, our findings overlap related to investing in awareness regarding neurodiversity, create a workplace where people with ADHD feel safe, and some common agile practices adaptations (e.g., joint estimation, agree definition of done, pair programming to fight procrastination). Specifically regarding Agile practices, we provide concrete suggestions for such adapted practices taking into consideration people with ASD as well. Concerning pair programming, we went further by understanding the need for familiarity with the team mate. In the case of Definition of Done for a task, we also expanded it to understand the need for clearer instructions and the importance of breaking down tasks.

\section{Threats and Limitations}
The sample size, although sufficient for a qualitative inquiry, limits the generalizability of the findings across all neurodivergent populations in the SE field. Also, relying on self-reported experiences may introduce bias, as participants may not fully recall or accurately convey facts about professional experiences related to their neurodivergent condition. In addition, the study predominantly focused on individuals with ADHD and ASD, excluding other neurodivergent conditions, which could have broadened the theoretical insights. An additional threat to validity is that two of the authors are neurodivergent themselves, potentially leading them to direct interviews in a certain direction or interpret the data based on their own journey way. To address this, we also included neurotypical authors in the interview guide construction and in analysis discussions, but also highlight that such threat would similarly exist in a purely neurotypical team. Finally, despite the interdisciplinary nature of this research, the findings are restricted to SE teams, potentially limiting the applicability of the results to other technical domains.

\section{Conclusions}
In this paper, we present a theory of how neurodivergent cognitive dysfunctions affect Software Engineering (SE) performance, and how this effect is moderated through the individual's journey and through accommodations. 
Our findings provide a valuable source to better understand how neurodivergent individuals -- especially with ADHD and ASD -- experience their workplace and SE tasks. Individuals with ASD+ADHD co-occurrence may struggle even more, since they have dysfunctions from both disorders. The theory lays a solid foundation for providing accommodations. As a part of these, and as the main novelty of our contribution, we provide empirically-grounded recommendations on how to tailor Agile practices to improve inclusion of neurodivergent engineers in SE teams.

\bibliographystyle{IEEEtran}
\bibliography{asd_adhd.bib}

\end{document}